# Redesign of Web-Based Exam for Knowledge Evaluation in Advanced Mathematics for Pharmaceutical Students Based on Analysis of the Results

Gergana Maneva[1, a)] and Mancho Manev[2, 3, b)]

Author Affiliations
[1]*University of Plovdiv Paisii Hilendarski, Faculty of Pedagogy, Department of Social Activities, 24 Tzar Asen St., Plovdiv 4000, Bulgaria.*
[2]*University of Plovdiv Paisii Hilendarski, Faculty of Mathematics and Informatics, Department of Algebra and Geometry, 24 Tzar Asen St., Plovdiv 4000, Bulgaria*
[3]*Medical University – Plovdiv, Faculty of Pharmacy, Department of Medical Physics and Biophysics, 15A Vasil Aprilov Blvd., Plovdiv 4002, Bulgaria*

*Author Emails*
[a)]gmaneva@uni-plovdiv.bg
[b)] Corresponding author: mmanev@uni-plovdiv.bg, mancho.manev@mu-plovdiv.bg

**Abstract.** The usage of the information technologies everywhere leads to demands for new manners of education. Modern e-learning environments lead the teaching, the learning and the evaluation of acquired knowledge and skills of the students to a new era. The students' motivation for e-learning is considered. The course of Advanced Mathematics is part of the curriculum of pharmaceutical students at the author's university. For students' knowledge evaluation it is used a hybrid-type exam in this university discipline, i.e., a problems-solving part and a remote web-based test which is created using the free and open-source e-educational platform Moodle. This paper presents a detailed analysis of the implemented test for knowledge evaluation of the students, using statistical methods and instruments. The information from the test is estimated and analyzed. Thus, it is made an improvement of the database of the test questions and it is enhanced the quality of the developed knowledge evaluation.

## INTRODUCTION

The students' involvement in teaching-learning process leads to improvement of the effectiveness of the education. This involvement could be achieved using interactive courses. To help with this process the educators use modern web-based e-learning systems.

The improvement of the university discipline Advanced Mathematics for pharmaceutical students is a complex activity. Many factors determine the success of it, such as the quality of instructional materials, the pedagogic skills of the teachers, the presentation of the content, the educational environment, the motivation of the students and so on. All these factors are important parts of the teaching-learning process. They must be kept in view of any effort to ensure the quality of the modern university education. The use of information technologies in all spheres of life leads to demands for introduction new techniques into teaching-learning process.

The idea of proposing and completing tasks through innovative distance learning platforms is identified not only as necessary but also attractive. This is based on the psychological and pedagogical needs of the current generation of students.

The main reason for the need of an interactive intervention model is the essence of the new generation. Generation Z is made up of representatives who are often over-receptive. Strong and effective stimulation is needed to gain their attention. In-formation is best presented visually because they often lose interest in long text or words. Generation Z is materially adjusted. They cannot wait and just wants things happen quickly. Born in the World of Technology, not only they do not remember the time before the computers and smartphones, they can't even imagine it. This is extremely difficult for face-to-face communication. Independent, remote work is preferred [4].

The idea of a web-based teaching-learning model is to impart new knowledge or practice and to exploit old ones. At the same time, this process is related and de-pendent on the actions and awareness of the students. A significant feature of the web-based model is that as an active player in the process the students' motivation to work increases. This means "getting" students inside the work process is very important.

## THE UNIVERSITY DISCIPLINE ADVANCED MATHEMATICS

Such objectives of teaching Advanced Mathematics like critical and analytical thinking, logical reasoning, decision-making, problem-solving are very important and it is difficult to be achieved only through verbal and mechanical methods. So, in the educational process in the authors' university, it is realized a hybrid educational course in the university discipline Advanced Mathematics, consisting of attendance hours for lectures and seminars, combined with remote implementation of a theoretical test. Nowadays the usage of online tests as a students' knowledge and skills evaluation has increased significantly. Some of the main reasons for this are:

- high efficiency, i.e., for a small period of time the lecturers could assess a large number of students;
- opportunity for simultaneous evaluation of students from different specialties, courses and groups;
- the immediate student's assessment after completing the test;
- the teacher's ability to manage, expand and redesign the database of the tests' questions and the evaluation criteria.

In many universities for interactive educational process, it is mainly used the open-source system Moodle [5], [6], [8], [9], [13]. There are also many other university projects in authors country which are integrated into the studying process [7], [12]. Other works that lead to improvement of the effectiveness of the electronic education are [2], [10], [14], [15], [16].

**TABLE 1.** Two types of testing (Source: Author, 2017)

| Student | Paper-based test | | Electronic test | |
|---|---|---|---|---|
| | Percentage | Assessment | Percentage | Assessment |
| Student №1 | 60,00 | Very Good (4,50) | 60,00 | Very Good (4,50) |
| Student №2 | 60,00 | Very Good (4,50) | 73,33 | Excellent (5,50) |
| Student №3 | 70,00 | Excellent (5,50) | 63,33 | Very Good (4,50) |
| Student №4 | 80,00 | Excellent (6,00) | 80,00 | Excellent (6,00) |
| Student №5 | 63,33 | Very Good (4,50) | 63,33 | Very Good (4,50) |
| Student №6 | 63,33 | Very Good (4,50) | 70,00 | Excellent (5,50) |
| Student №7 | 80,00 | Excellent (6,00) | 90,00 | Excellent (6,00) |
| Student №8 | 80,00 | Excellent (6,00) | 80,00 | Excellent (6,00) |
| Student №9 | 73,33 | Excellent (5,50) | 73,33 | Excellent (5,50) |
| Student №10 | 73,33 | Excellent (6,00) | 83,33 | Excellent (6,00) |
| Average | 70,33 – Excellent (5,50) | | 73,67 – Excellent (5,50) | |

In [8], using the e-education system Moodle we developed a methodological approach of a hybrid educational course, i.e., a compulsory attendance at lectures and seminars in combination with two different types of conduction of the final test for comparison – a paper-based test and a remote web-based one. In the cited work, there are implemented to the one and the same group of 10 students the electronic test and a paper-based one for knowledge assessment. The evaluation is performed by a percentage scale (0-100%) corresponding to estimates by five-point scale: Fail (2) – Satisfactory (3) – Good (4) – Very good (5) – Excellent (6) (Table 1).

The results received from the parallel tests and the positive feedback from students led to the conclusion for the continuation of this form of teaching in the future.

## REDESIGN OF THE INTERACTIVE COURSE

Every newly created test needs improvement. This improvement means to add new questions to the database, to correct and refine existing ones. Now, we show the procedure of redesigning of our educational course.

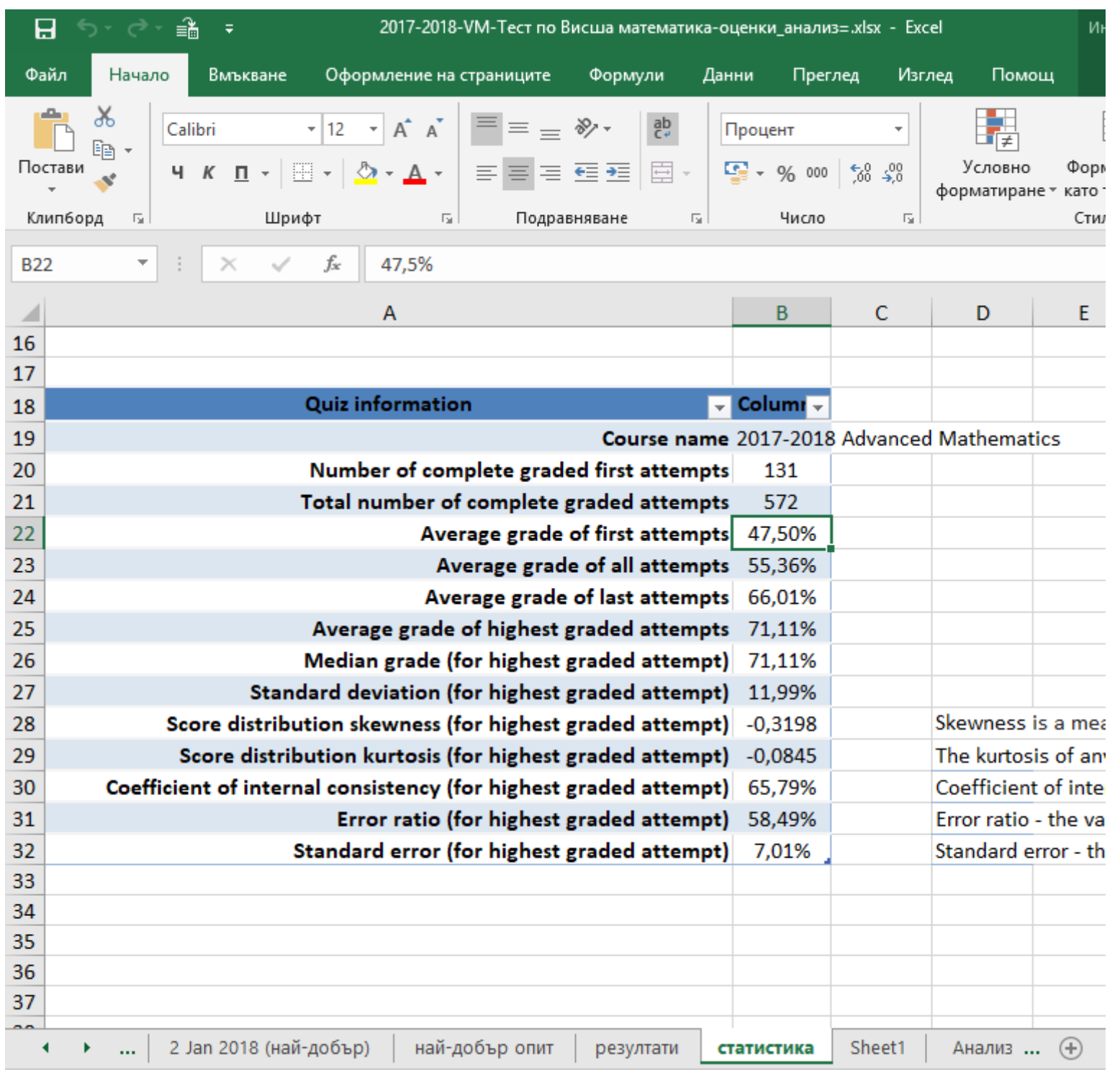

| Quiz information | Column | |
|---|---|---|
| Course name | 2017-2018 Advanced Mathematics | |
| Number of complete graded first attempts | 131 | |
| Total number of complete graded attempts | 572 | |
| Average grade of first attempts | 47,50% | |
| Average grade of all attempts | 55,36% | |
| Average grade of last attempts | 66,01% | |
| Average grade of highest graded attempts | 71,11% | |
| Median grade (for highest graded attempt) | 71,11% | |
| Standard deviation (for highest graded attempt) | 11,99% | |
| Score distribution skewness (for highest graded attempt) | -0,3198 | Skewness is a mea |
| Score distribution kurtosis (for highest graded attempt) | -0,0845 | The kurtosis of an |
| Coefficient of internal consistency (for highest graded attempt) | 65,79% | Coefficient of inte |
| Error ratio (for highest graded attempt) | 58,49% | Error ratio - the va |
| Standard error (for highest graded attempt) | 7,01% | Standard error - th |

**FIGURE 1.** Test statistics (Source: Authors)

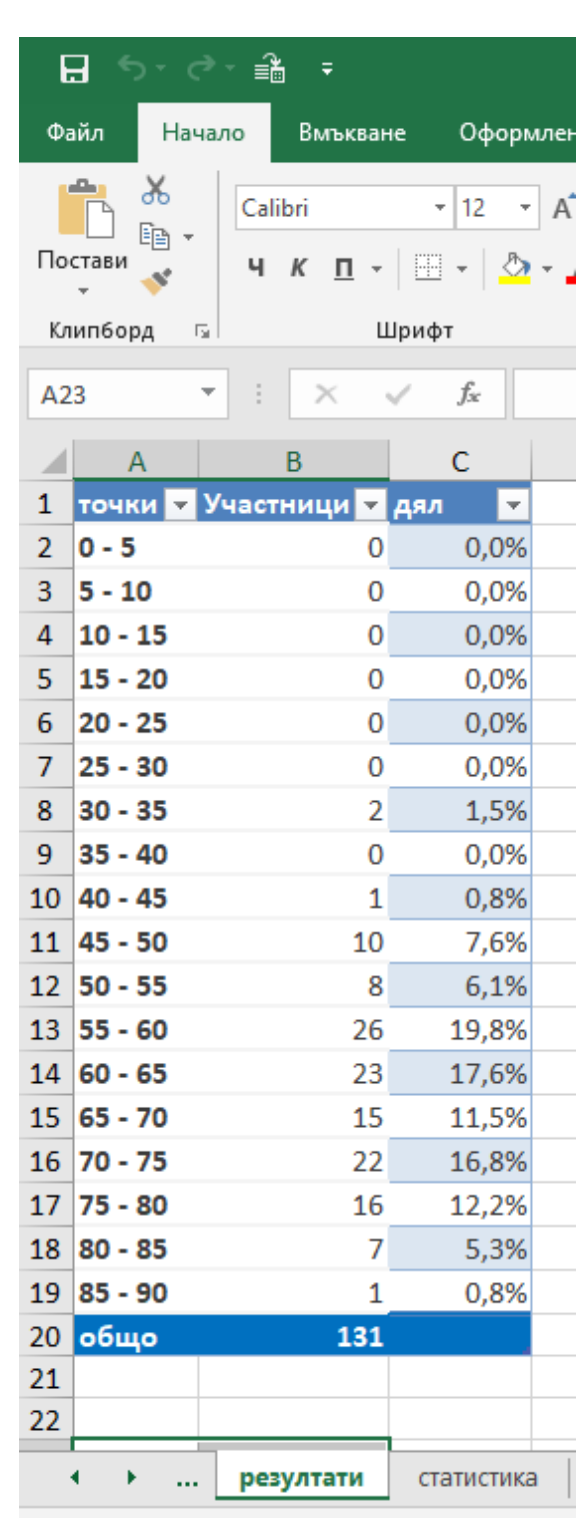

| точки | Участници | дял |
|---|---|---|
| 0 - 5 | 0 | 0,0% |
| 5 - 10 | 0 | 0,0% |
| 10 - 15 | 0 | 0,0% |
| 15 - 20 | 0 | 0,0% |
| 20 - 25 | 0 | 0,0% |
| 25 - 30 | 0 | 0,0% |
| 30 - 35 | 2 | 1,5% |
| 35 - 40 | 0 | 0,0% |
| 40 - 45 | 1 | 0,8% |
| 45 - 50 | 10 | 7,6% |
| 50 - 55 | 8 | 6,1% |
| 55 - 60 | 26 | 19,8% |
| 60 - 65 | 23 | 17,6% |
| 65 - 70 | 15 | 11,5% |
| 70 - 75 | 22 | 16,8% |
| 75 - 80 | 16 | 12,2% |
| 80 - 85 | 7 | 5,3% |
| 85 - 90 | 1 | 0,8% |
| общо | 131 | |

**FIGURE 2.** Distribution of the marks (Source: Authors)

A great practical value for the development of an interactive Moodle course has the submenu "Statistics" (Administration > Quiz administration > Results > Statistics). It provides a statistical analysis of the test and the questions within it. A dropdown menu allows the lecturer to select which attempts of the students to be involved in the calculation of the statistics. Also, it could be chosen whether to display information about all attempts or just the first attempt. The Moodle system allows also more detailed analysis of each question individually. The full text of the statistic report can be downloaded in various formats. The most useful one of them is in Excel spreadsheet file. The report gives information about the number of evaluated students' attempts, the average assessments of the first, last attempt and overall, the median of the evaluations and other useful statistical values (Fig. 1). Moreover, the report shows the distribution of the marks (Fig. 2). In Fig. 3, it is shown a statistical report that is used for several reasons.

We find the so called "broken" questions. In the first columns are given details about the question and the number of attempts. A useful tool here is the so called "Facility index (F)". This index shows the mean score of students on the chosen item. Table 2 shows the interpretations of the respective F.

The next column is devoted to the "Standard deviation" of each question. This index measures the spread of scores about the mean and hence the extent to which the question might discriminate. If the Facility index is very high or very low it is impossible the spread to be large. A value of Standard deviation less than 33% in the table is not generally satisfactory.

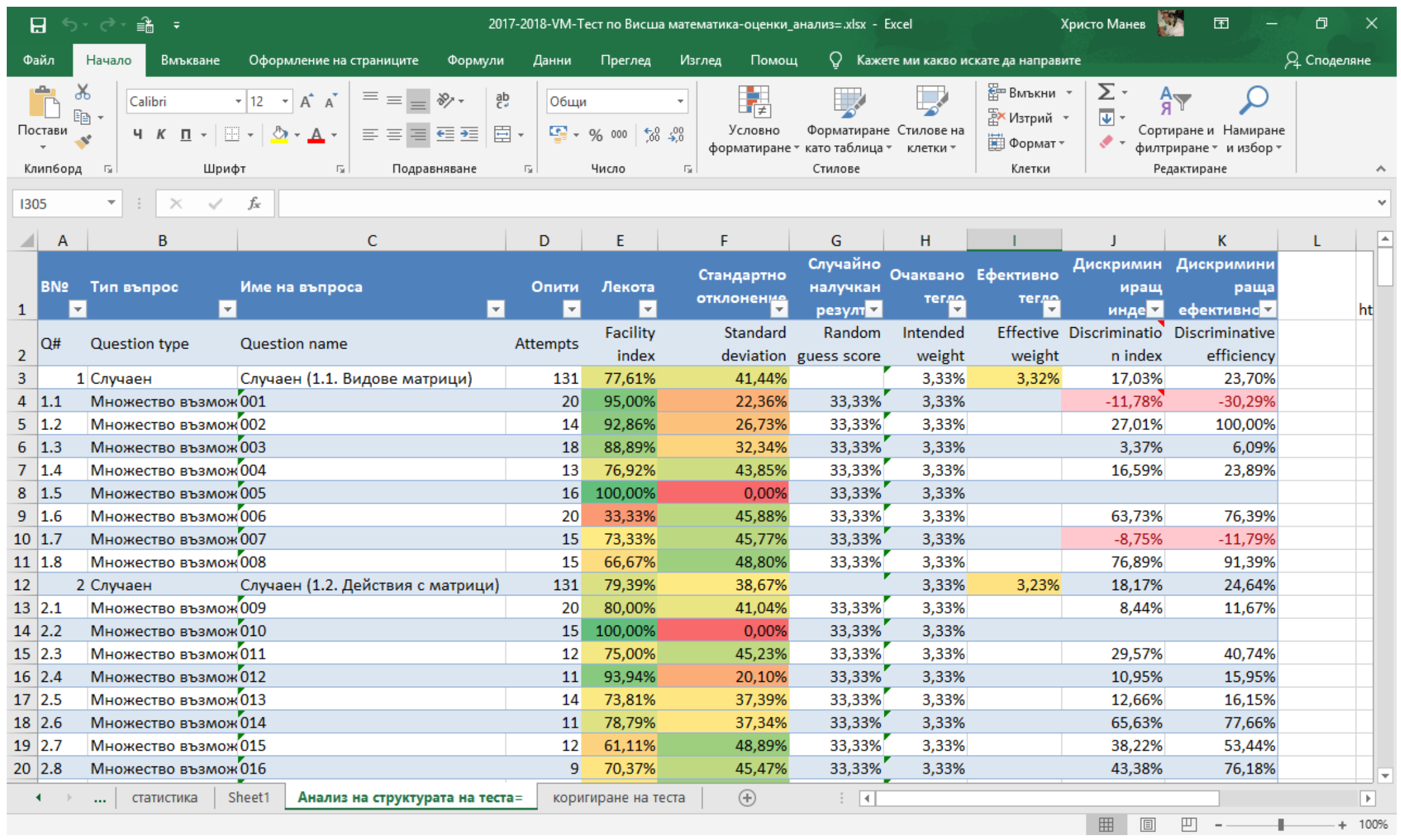

| BNº | Тип въпрос | Име на въпроса | Опити | Лекота | Стандартно отклонение | Случайно налучкан резулт | Очаквано тегло | Ефективно тегло | Дискриминиращ инде | Дискриминираща ефективно |
|---|---|---|---|---|---|---|---|---|---|---|
| Q# | Question type | Question name | Attempts | Facility index | Standard deviation | Random guess score | Intended weight | Effective weight | Discrimination index | Discriminative efficiency |
| 1 | Случаен | Случаен (1.1. Видове матрици) | 131 | 77,61% | 41,44% | | 3,33% | 3,32% | 17,03% | 23,70% |
| 1.1 | Множество възмож | 001 | 20 | 95,00% | 22,36% | 33,33% | 3,33% | | -11,78% | -30,29% |
| 1.2 | Множество възмож | 002 | 14 | 92,86% | 26,73% | 33,33% | 3,33% | | 27,01% | 100,00% |
| 1.3 | Множество възмож | 003 | 18 | 88,89% | 32,34% | 33,33% | 3,33% | | 3,37% | 6,09% |
| 1.4 | Множество възмож | 004 | 13 | 76,92% | 43,85% | 33,33% | 3,33% | | 16,59% | 23,89% |
| 1.5 | Множество възмож | 005 | 16 | 100,00% | 0,00% | 33,33% | 3,33% | | | |
| 1.6 | Множество възмож | 006 | 20 | 33,33% | 45,88% | 33,33% | 3,33% | | 63,73% | 76,39% |
| 1.7 | Множество възмож | 007 | 15 | 73,33% | 45,77% | 33,33% | 3,33% | | -8,75% | -11,79% |
| 1.8 | Множество възмож | 008 | 15 | 66,67% | 48,80% | 33,33% | 3,33% | | 76,89% | 91,39% |
| 2 | Случаен | Случаен (1.2. Действия с матрици) | 131 | 79,39% | 38,67% | | 3,33% | 3,23% | 18,17% | 24,64% |
| 2.1 | Множество възмож | 009 | 20 | 80,00% | 41,04% | 33,33% | 3,33% | | 8,44% | 11,67% |
| 2.2 | Множество възмож | 010 | 15 | 100,00% | 0,00% | 33,33% | 3,33% | | | |
| 2.3 | Множество възмож | 011 | 12 | 75,00% | 45,23% | 33,33% | 3,33% | | 29,57% | 40,74% |
| 2.4 | Множество възмож | 012 | 11 | 93,94% | 20,10% | 33,33% | 3,33% | | 10,95% | 15,95% |
| 2.5 | Множество възмож | 013 | 14 | 73,81% | 37,39% | 33,33% | 3,33% | | 12,66% | 16,15% |
| 2.6 | Множество възмож | 014 | 11 | 78,79% | 37,34% | 33,33% | 3,33% | | 65,63% | 77,66% |
| 2.7 | Множество възмож | 015 | 12 | 61,11% | 48,89% | 33,33% | 3,33% | | 38,22% | 53,44% |
| 2.8 | Множество възмож | 016 | 9 | 70,37% | 45,47% | 33,33% | 3,33% | | 43,38% | 76,18% |

**FIGURE 3.** Analysis of the test – Statistical indices (Source: Authors)

"Random guess score" is the next column which is the mean score students would be expected to get for a random guess at the question. The next two statistical values are "Intended weight" and "Effective weight". The first one shows the question weight expressed as a percentage of the overall test score and the other one estimates the weight the question actually has in contributing to the overall spread of scores. If the effective weight is greater than the intended weight it shows the question has a greater share in the spread of scores than may have been intended. If it is less than the intended weight it shows that it is not having as much effect in spreading out the scores as was intended.

**TABLE 2.** Interpretations of F (Source: Moodle)

| *F* | Interpretation |
|---|---|
| 5 or less | Extremely difficult or something wrong with the question |
| 6 – 10 | Very difficult |
| 11 – 20 | Difficult |
| 21 – 34 | Moderately difficult |
| 36 – 65 | About right for the average student |
| 66 – 80 | Fairly easy |
| 81 – 89 | Easy |
| 90 – 94 | Very easy |
| 95 – 100 | Extremely easy |

The next analysis feature shows up in the "Discrimination index" column. This is the correlation between the weighted scores on the question and those on the rest of the test. It indicates how effective the question is at sorting out able students from those who are less able. If the number there is small, you may have a problem. Moodle is highlighting any low values in this column so they stand out. The results (R) should be interpreted in Table 3 as follows:

The last column here is "Discrimination efficiency". This statistic feature estimates how good the discrimination index is relative to the difficulty of the question. The discrimination efficiency will very rarely approach 100%, but values in excess of 50% should be achievable. Lower values indicate that the question is not nearly as effective at discriminating between students of different ability as it might be and therefore is not a particularly good question.

**TABLE 3.** Interpretations of R (Source: Moodle)

| *R* | Discrimination |
|---|---|
| 50 or more | Very good |
| 30 – 50 | Adequate |
| 20 – 30 | Weak |
| 0 – 20 | Very weak |
| 0 or less | Probably invalid question |

Also, we understand how students are responding to a particular question. If the lecturer clicks through to the details about a particular question, then at the bottom of that page are shown all the different responses that were given, whether they were marked right or wrong, and how many students gave each result. This is used by us to notice a potential problem (Fig. 4).

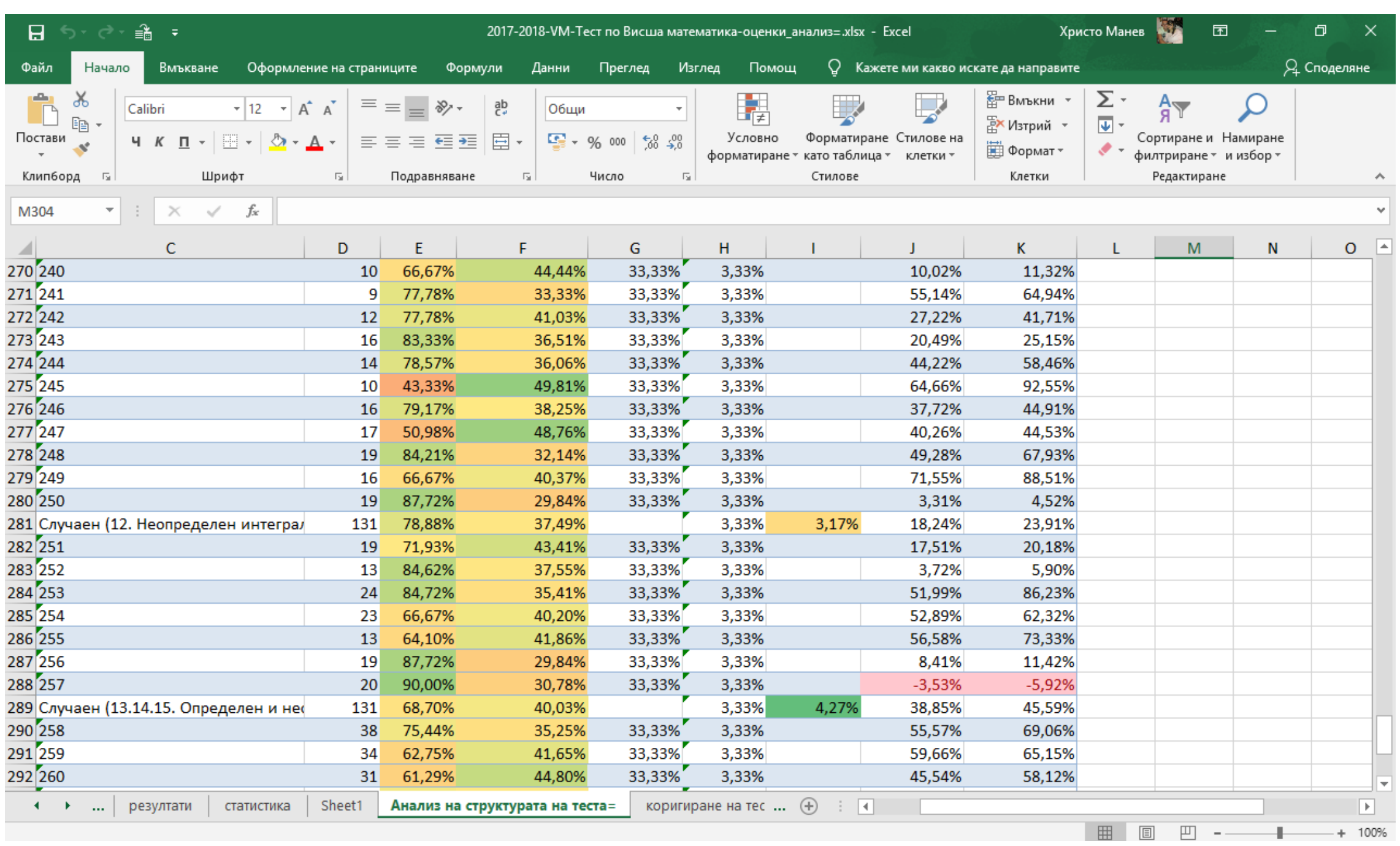

| | C | D | E | F | G | H | I | J | K |
|---|---|---|---|---|---|---|---|---|---|
| 270 | 240 | 10 | 66,67% | 44,44% | 33,33% | 3,33% | | 10,02% | 11,32% |
| 271 | 241 | 9 | 77,78% | 33,33% | 33,33% | 3,33% | | 55,14% | 64,94% |
| 272 | 242 | 12 | 77,78% | 41,03% | 33,33% | 3,33% | | 27,22% | 41,71% |
| 273 | 243 | 16 | 83,33% | 36,51% | 33,33% | 3,33% | | 20,49% | 25,15% |
| 274 | 244 | 14 | 78,57% | 36,06% | 33,33% | 3,33% | | 44,22% | 58,46% |
| 275 | 245 | 10 | 43,33% | 49,81% | 33,33% | 3,33% | | 64,66% | 92,55% |
| 276 | 246 | 16 | 79,17% | 38,25% | 33,33% | 3,33% | | 37,72% | 44,91% |
| 277 | 247 | 17 | 50,98% | 48,76% | 33,33% | 3,33% | | 40,26% | 44,53% |
| 278 | 248 | 19 | 84,21% | 32,14% | 33,33% | 3,33% | | 49,28% | 67,93% |
| 279 | 249 | 16 | 66,67% | 40,37% | 33,33% | 3,33% | | 71,55% | 88,51% |
| 280 | 250 | 19 | 87,72% | 29,84% | 33,33% | 3,33% | | 3,31% | 4,52% |
| 281 | Случаен (12. Неопределен интеграл | 131 | 78,88% | 37,49% | | 3,33% | 3,17% | 18,24% | 23,91% |
| 282 | 251 | 19 | 71,93% | 43,41% | 33,33% | 3,33% | | 17,51% | 20,18% |
| 283 | 252 | 13 | 84,62% | 37,55% | 33,33% | 3,33% | | 3,72% | 5,90% |
| 284 | 253 | 24 | 84,72% | 35,41% | 33,33% | 3,33% | | 51,99% | 86,23% |
| 285 | 254 | 23 | 66,67% | 40,20% | 33,33% | 3,33% | | 52,89% | 62,32% |
| 286 | 255 | 13 | 64,10% | 41,86% | 33,33% | 3,33% | | 56,58% | 73,33% |
| 287 | 256 | 19 | 87,72% | 29,84% | 33,33% | 3,33% | | 8,41% | 11,42% |
| 288 | 257 | 20 | 90,00% | 30,78% | 33,33% | 3,33% | | -3,53% | -5,92% |
| 289 | Случаен (13.14.15. Определен и нес | 131 | 68,70% | 40,03% | | 3,33% | 4,27% | 38,85% | 45,59% |
| 290 | 258 | 38 | 75,44% | 35,25% | 33,33% | 3,33% | | 55,57% | 69,06% |
| 291 | 259 | 34 | 62,75% | 41,65% | 33,33% | 3,33% | | 59,66% | 65,15% |
| 292 | 260 | 31 | 61,29% | 44,80% | 33,33% | 3,33% | | 45,54% | 58,12% |

**FIGURE 4.** Analysis of the test - responds to a particular question (Source: Authors)

The listed tools helped us to correct all the inaccuracies and to mark and redesign all the questions that are correct but are not set in a very clear habit (Fig. 5). We find different types of necessity to correct a question, for example:

- Question 110 – The question and possible answers were formulated correctly but many wrong answers were given. Obviously, the students need more time to clarify this part of the study material;
- Question 95 – Incorrectly formulated possible answers;
- Question 64 – Not very well worded wrong answers;
- Question 74 – There is a need for refinement of mathematical terminology.

All these inaccuracies and errors were corrected. We plan to test the upgraded version in a similar way and to make a comparative analysis of the two results.

Often speaking of web-based learning models, the main focus is on the efficiency of the knowledge transmitted or the successful achievement of the required educational standards by students. It is about the quality and effectiveness of online learning. Models are being sought to increase and improve them. However, the question remains about students' motivation to learn. What happens to it when moving from standard learning methods to web-based learning models?

| | A | B | C | D | E | F | G | H | I | J | K | L | M | N | O | P | Q |
|---|---|---|---|---|---|---|---|---|---|---|---|---|---|---|---|---|---|
| 19 | | | | върху една равнина | | | | | | | | | | | | | |
| 20 | | | | | | | | | | | | | | | | | |
| 21 | 9.12 | Множеств | 076 | 11 | 18,18% | 31,14% | 33,33% | 3,33% | | -12,63% | -18,79% | | | | | | |
| 22 | | | Един вектор **a** е линейно зависим тогава и само тогава, когато: | | | | | | | Множество от един вектор **a** е линейно зависима система вектори тогава и само тогава, когато: | | | | | | | |
| 23 | | | | **a** не е нулев | | | | | | грешно формулиран въпрос за отговора | | | | | | | |
| 24 | | | | **a** е нулев | | | | | | | | | | | | | |
| 25 | | | | съществува вектор **b**, така че **b** = λ**a** за някакво число λ | | | | | | | | | | | | | |
| 26 | | | | | | | | | | | | | | | | | |
| 27 | 9.17 | Множеств | 081 | 8 | 25,00% | 15,43% | 33,33% | 3,33% | | -27,09% | -38,71% | | | | | | |
| 28 | | | Ако **a** и **b** са неколинеарни вектори, то линейно независими са следните три вектора: | | | | | | | *Ако a и b са неколинеарни вектори, то системата от трите вектора {a, b, c} е линейно независим* | | | | | | | |
| 29 | | | | **a**, **b** и произволна тяхна линейна комбинация | | | | | | | линейна комбинация на **a** и **b** | | | | коректен въпрос и отговори | | |
| 30 | | | | **a**, **b** и тяхното векторно произведение | | | | | | | векторното произведение на **a** и **b** | | | | вероятно неясен | | |
| 31 | | | | **a**, **b** и нулевият вектор | | | | | | | нулевият вектор | | | | | | |
| 32 | | | | | | | | | | | | | | | | | |
| 33 | 13.2 | Множеств | 095 | 18 | 22,22% | 37,92% | 33,33% | 3,33% | | 11,83% | 15,98% | | | | | | |
| 34 | | | Ако е дадено условието **ab** = 0 за скаларното произведение, то следва, че: | | | | | | | Ако е дадено условието **ab** = 0 за скаларното произведение, то следва, че: | | | | | | | |
| 35 | | | | или **a**, или **b** е нулев вектор | | | | | | | или **a**, или **b** е нулев вектор | | | | грешно формулирани отговори | | |
| 36 | | | | **a** и **b** са ненулеви и взаимно ортогонални вектори | | | | | | | **a** и **b** са или нулеви или линейно зависими вектори | | | | | | |
| 37 | | | | **a** и **b** са ненулеви или взаимно ортогонални вектори | | | | | | | ако **a** и **b** са ненулеви, то те са взаимно ортогонални вектори | | | | | | |
| 38 | | | | | | | | | | | | | | | | | |
| 39 | 13.3 | Множеств | 096 | 12 | 25,00% | 37,94% | 33,33% | 3,33% | | -4,50% | -4,97% | | | | | | |
| 40 | | | Скаларното произведение притежава свойството: | | | | | | | | | | | | | | |
| 41 | | | | $(a - b)_2 = a_2 - 2ab + b_2$ | | | | | | | | | | | вярно е всичко | | |

**FIGURE 5.** Redesign of the test (Source: Authors)

In [1], it is made a study that examines motivating and demotivating factors of the usage of Moodle. The survey encompasses data collected from 157 respondents (Russia) with the focus on students' and teachers' perception of Moodle-based learning. The conclusion is reached that technical support has a direct effect on the perceived ease of use and perceived usefulness. According to the data collected from junior students, the overwhelming majority of them (92 %) agreed on the fact that an opportunity to get extra points to their attestation encourage them to use Moodle. The survey results also indicate that receiving feedback from the teacher via Moodle increases the learners' motivation as evidenced. The students reported that their motivation is also based on the way the online learning material is organized and the opportunity for collaborative learning was regarded as motivating. The most demotivating factor given by the respondents is the technical problems while working with the electronic platform. The results show that the most attractive for teachers is the possibility for automatic verification, the free sharing of additional information, as well as the possibility for an individual approach. This may summarize that the Moodle platform provides a field of thought that does not unnecessarily simplify the work process.

In [11], the author explores how the use of Moodle in the university process can increase students' motivation. The survey uses a questionnaire on 52 students (Romania). The results show that 88% of respondents want to use the Moodle platform more often. This leads to the conclusion that the virtual learning environment increases students' motivation and facilitates the learning process. The majority of participants (57%) believe that the platform increases the effectiveness of the learning process. The study highlights some of Moodle's strengths: a virtual library with unlimited access, easy communication and teamwork, and a direct individual connection with the professor. Extremely interesting is the fact that according to the surveyed students' online assessment is much more objective than traditional.

A study in Turkey found that Moodle contributes significantly to students' levels of motivation and autonomy [3]. The author also chooses a specific definition of what is motivation in learning: "development of mental processes, learning outcomes and behavior". The difference between traditional motivation to learn and motivation in an online environment is emphasized. All synchronous or asynchronous elements affect the cognitive processes associated with learning motivation. According to the author, one of the main advantages of the Moodle platform is the ability to build a stable and trusting relationship and communication between students and teachers. The article offers an interesting look at the possible challenges of using Moodle. They are divided into three categories:

- challenges for professors - time and system management;
- challenges for the institution - internet access, active library content, online assessment;
- challenges for students - student trends, plagiarism, student profiles.

The results show that the usage of the e-learning platform Moodle inspires and increases students' motivation and desire to use a similar type of teaching-learning process in the future. The use of the Moodle platform provides learning autonomy. The work once again emphasizes the needs of the new generation.

## RESULTS AND CONCLUSION

This work presents a detailed analysis of the implemented electronic test for knowledge evaluation of the students, using statistical methods, instruments and the opportunities provided by the educational e-platform Moodle. The questions included in the test and the respective answers given by the students were estimated and analyzed. Thus, it is made an improvement of the database of the test questions. The inaccuracies were corrected, the wording of some of the questions which were not well understood by students was improved and the way of selecting questions from different themes was fine-tuned. The received results are used to enhance the quality of the developed knowledge evaluation and the type of its implementation.

Based on the findings of this paper, the positives of working with web-based platforms can be confirmed. The theoretical overview of the topic and the concrete results presented in this work show on one more time the need for scientific and theoretical development in this field.

It is important to make a recommendation for future development on the topic that logically develops the goals and objectives of the study. For further developing Moodle's abilities and empirically testing their effectiveness, it is important to pay attention to students' opinions and attitudes about this method of learning. A recommendation is made for the future development of students' feedback. The present study provides the necessary guidelines for compiling the required survey:

- To examine respondents' age characteristics and their attitude to the web-based teaching method.
- Find or reject the existence of gender differentiation in students' opinions and attitudes about the Moodle platform.
- To look for correlation between the studied discipline and the respondents' attitudes towards using web-based teaching methods.
- To investigate possible changes in the levels of motivation to learn in the proposed innovative method compared to the standard work process.

The results achieved and the conclusions drawn in this work should serve as a basis for further in-depth research in the field. It is necessary to deepen the study of the effectiveness of the proposed method and what is the reason of extension of the respondents' better achievements as a result of it. It would be neat to statistically extent also the comparison between the e-learning results and the results given by using the standard teaching and testing method. Moreover, a method for expanding the study and its interest in it would also be a comparison between the education results achieved by dint of the Moodle system and those achieved through other online distance learning platforms.

It is expected that the creation of a feedback survey, the empirical verification of the respondents' opinion, the exploration of the platform's capabilities and its competitiveness with similar ones, will enable the specialists to develop more the means for effective and efficient online training. This, in turn, will lead to the deepening of specific competences and the professional development of the educators. This evolutionary process is likely to continue in the future as the unique experience and knowledge of professionals – all so-called e-Teachers become more widespread and important in the field of education.

In the digital age, it is imperative that we all take advantage of the opportunities that are available to us. Undoubtedly, online learning gives unparalleled freedom to both students and teachers. The ability for everyone to do the job themselves at convenient time and place results in providing a much easier and dynamic flow of information.

The successful integration of e-learning, the better interaction with the learners and the shown good results and positive feedback are preconditions for the continuation of this form of teaching in the future. Moreover, the future examined students in this course will fill in an anonymous poll to show the authors their thoughts for this type of hybrid educational system and the given responses will be taken into account in the future development of this type of education.

## ACKNOWLEDGMENTS

The second author was partially supported by Medical University – Plovdiv and Project FP21-FMI-002 of the Scientific Research Fund, University of Plovdiv.

## REFERENCES

1. T. Y. Aikina, and L. M. Bolsunovskaya, Int. J. Emerg. Technol. Learn. 15 (2), 239–248 (2020).
2. D. Alyahya and N. Almutairi, Int. Educ. Stud. **12** (5), 109–119 (2019).
3. E. Ayan, Open J. Mod. Linguist. **5**, 6–20 (2015).
4. E. J. Cilliers, Int. J. Soc. Sci. **3** (1), 188–198 (2017).
5. J. R. Cole, *Using Moodle: Teaching with the Popular Open Source Course Management System* (O'Reilly Community Press, Sebastopol, 2007).
6. H. Manev and S. Enkov, Interactive education in Graph Theory, In: Proceedings of Fifth National Conference EIS, Plovdiv, 41-49 (2012).
7. H. Manev and A. Golev, "Electronic textbook "Geometry for students in Informatics" with DisPeL," in *Proceedings of International Conference "From DeLC to Velspace"*, (Third Millennium Media Publications, Plovdiv, 2014), pp. 191–198.
8. H. Manev and M. Manev, "Design, analysis and implementation of electronic test for knowledge evaluation in the course of Information Technologies for pharmaceutical students," in *CBU International Conference Proceedings: Innovations in Science and Education* **5** (CBU Research Institute, Praha, 2017), pp. 705–709.
9. H. Manev and T. Terzieva, Comput. Sci. Educ. Comput. Sci. **11** (1), 78–84 (2015).
10. C. M. Millsap, "Comparison of computer testing versus traditional paper-and-pencil testing," Ph.D. thesis, University of North Texas, 2000.
11. G. C. Oproiu, Procedia Soc. Behav. Sci. **180**, 426–432 (2015).
12. A. Rahnev, N. Pavlov and V. Kyurkchiev, Eur. Int. J. Sci. Technol. **3** (1), 95–109 (2014).
13. W. Rice, *Moodle E-Learning Course Development: A complete guide to successful learning using Moodle* (Packt Publishing, Birmingham, 2006).
14. M. R. Simonson, M. Maurer, M. Montag-Torardi and M. Whitaker, J. Educ. Comput. Res. **3**, 231–247 (1987).
15. T. J. Ward, S. R. Hooper and K. M. Hannafin, J. Educ. Comput. Res. **5**, 327–333 (1989).
16. D. Weiss and G. Kingsbury, J. Educ. Meas. **21**, 361–375 (1984).